\documentclass[apj,twocolumn,a4paper,final]{aastex631}
\usepackage[utf8]{inputenc}
\usepackage[T1]{fontenc}
\usepackage{amsmath, amssymb}
\usepackage{graphicx}
\usepackage[svgnames,dvipsnames]{xcolor}
\usepackage{booktabs}
\usepackage{natbib}
\usepackage{ifdraft}
\usepackage{hyperref}
\usepackage{savesym}
\savesymbol{tablenum}
\usepackage[
    free-standing-units,
    exponent-product = \times,
    list-units = single,
    range-units = single,
    multi-part-units = single,
    separate-uncertainty = true,
]{siunitx}
\DeclareSIUnit{\sigmaunit}{\ensuremath{\sigma}}
\restoresymbol{SIX}{tablenum}
\AtBeginDocument{\RenewCommandCopy\qty\SI}
\usepackage{acro} %
\usepackage{csquotes} %
\usepackage[capitalize]{cleveref}
\usepackage{etoolbox}
\graphicspath{{./Figures}}
\makeatletter
\patchcmd{\ltx@foottext}{%
  .5\textwidth\advance\hsize-18pt}{%
  \linewidth\advance\hsize-1.8em%
}{}{}
\makeatother
\makeatletter
\pretocmd\@sect{\def\@currentcounter{#1}}{}{\fail}
\makeatother
\crefname{section}{Sec.}{Secs}

\definecolor{DipoleColor}{HTML}{DC143C}

\DeclareAcronym{gr}{
    short = GR ,
    long = general relativity
}

\DeclareAcronym{gw}{
    short = GW ,
    long = gravitational wave
}

\DeclareAcronym{bbh}{
    short = BBH ,
    long = binary black hole
}

\DeclareAcronym{grb}{
    short = GRB ,
    long = gamma-ray burst
}

\DeclareAcronym{flrw}{
    short = FLRW ,
    long = Friedmann-Lemaître-Robertson-Walker
}

\DeclareAcronym{gwtc}{
    short = GWTC ,
    long = Gravitational-Wave Transient Catalog 
}

\DeclareAcronym{lvk}{
    short = LVK ,
    long = LIGO-Virgo-KAGRA
}

\DeclareAcronym{sne}{
    short = SNe ,
    long = supernovae
}

\DeclareAcronym{snr}{
    short = SNR ,
    long = signal-to-noise ratio
}

\newcommand{\ie}{\textit{i.e}}
\newcommand{\eg}{\textit{e.g}}

\begin{document}
\title{Testing cosmological isotropy with gravitational waves and gamma-ray bursts}
\author{Brian H. Y. Cheng}
\affiliation{Department of Physics, The Chinese University of Hong Kong, Shatin, New Territories, Hong Kong}
\author{Donniel C. Cruz}
\affiliation{Department of Physics, De La Salle University, Manila, Philippines}
\author{Otto A. Hannuksela}
\affiliation{Department of Physics, The Chinese University of Hong Kong, Shatin, New Territories, Hong Kong}
\author{Davendra S. Hassan}
\affiliation{Department of Physics, National University of Singapore, Singapore 117551, Singapore}
\author{Christian Heiderijk}
\affiliation{Department of Physics, City University of Hong Kong, Kowloon, Hong Kong}
\author{Leo Q. Hu}
\affiliation{Department of Physics, City University of Hong Kong, Kowloon, Hong Kong}
\author{Souvik Jana}
\affiliation{Department of Physics, The Chinese University of Hong Kong, Shatin, New Territories, Hong Kong}
\author{Jinwon Kim}
\affiliation{Department of Physics, Chung-Ang University, Seoul, Korea}
\author{Albert K.~H.~Kong}
\affiliation{Institute of Astronomy, National Tsing Hua University, Hsinchu 30013, Taiwan}
\affiliation{Institute of Space Engineering, National Tsing Hua University, Hsinchu 30013, Taiwan}
\author{Peony K. K. Lai}
\affiliation{Department of Physics, The Chinese University of Hong Kong, Shatin, New Territories, Hong Kong}
\author{Samuel C. Lange}
\affiliation{Department of Physics, The Chinese University of Hong Kong, Shatin, New Territories, Hong Kong}
\author{Samson H. W. Leong}
\affiliation{Department of Physics, The Chinese University of Hong Kong, Shatin, New Territories, Hong Kong}
\author{Matteo Lulli}
\affiliation{Department of Physics, The Chinese University of Hong Kong, Shatin, New Territories, Hong Kong}
\author{Li-Ting Ma}
\affiliation{Institute of Astronomy, National Tsing Hua University, Hsinchu 30013, Taiwan}
\author{Paul Martens}
\affiliation{Department of Physics, The Chinese University of Hong Kong, Shatin, New Territories, Hong Kong}
\author{Boris H.-L. Ng}
\affiliation{Department of Physics, The Chinese University of Hong Kong, Shatin, New Territories, Hong Kong}
\author{Thomas C.~K.~Ng}
\affiliation{Nikhef, Science Park 105, 1098 XG Amsterdam, The Netherlands}
\affiliation{Institute for Gravitational and Subatomic Physics (GRASP), Utrecht University, Princetonplein 1, 3584 CC Utrecht, The Netherlands}
\affiliation{Department of Physics, The Chinese University of Hong Kong, Shatin, New Territories, Hong Kong}
\author{Surojit Saha}
\affiliation{Institute of Astronomy, National Tsing Hua University, Hsinchu 30013, Taiwan}
\author{Gwangeon Seong}
\affiliation{Department of Physics, Ewha Womans University, Seoul 03760, Korea}
\author{Helen Xian}
\affiliation{Department of Physics, The Chinese University of Hong Kong, Shatin, New Territories, Hong Kong}
\author{Yanyan Zheng}
\affiliation{Missouri University of Science and Technology, Rolla, MO 65409, USA }
\correspondingauthor{Paul Martens}
\email{paulmartens@cuhk.edu.hk}

\begin{abstract}
\noindent
The cosmological principle asserts that the Universe is homogeneous and isotropic on large enough scales.
However, alternative cosmological models can bring about anisotropies through local inhomogeneities, anisotropic evolution, or exotic physics.
In addition, select studies have also hinted at mild evidence of anisotropies in SNe Ia, CMB, and GRB data, though these remain unconfirmed.
In this work, we test for cosmological anisotropies using gravitational waves and gamma-ray bursts, adopting the latest O4a release from the LIGO-Virgo-KAGRA collaboration and GRBWeb (including all known GRBs since 1991).
If the cosmological principle holds, the sky localisation and the characteristics of the GRBs and GWs (masses, luminosities, redshifts) should be statistically isotropic when corrected for selection biases.
We employ a couple statistical methods, including angular power spectra and two-point correlation functions, and compare the results against synthetic data.
The work extends previous analyses by including the most recent datasets, and the use of multiple complementary statistical tests.
We find no significant evidence for anisotropy in the current GW and GRB datasets, consistent with the
cosmological principle.
\end{abstract}


\section{Introduction}
\label{sec:intro}

The cosmological principle states that the Universe is homogeneous and isotropic on large scales.
Most of modern cosmology relies on this assumption, largely via the \ac{flrw} metric, the most general metric under the principle's validity.
Observations revealed a dipole anisotropy pattern attributed to the kinematic effect of our motion relative to the cosmic rest frame.
The Planck collaboration~\citep{Planck:2019evm} measured a peculiar velocity of \qty{369.82(11)}{\kilo\meter\per\second} toward $(l, b) = (\qty{264.021(0.011)}{\degree},\ \qty{48.243(0.005)}{\degree})$ in galactic coordinates, \ie. $(\alpha, \delta) = (\ang{167.9},\ang{-6.9})$ in equatorial coordinate.
This kinematic dipole is expected to manifest itself in any cosmological tracer (\eg. galaxies, SNe, \acp{grb}, \acp{gw})~\cite{Ellis:1984uka}; though some authors have hinted at potential inconsistencies in amplitude with the CMB in a subset of tracers~\cite[e.g.][]{Secrest_2021,Secrest_2022,Land-Strykowski:2025gkz,Rasouli:2025yfv,Secrest:2025wyu,kinematic_dipole_4}, others have found results consistent with both the direction and amplitude of the CMB dipole~\cite{kinematic_dipole_favor_1, kinematic_dipole_favor_2}. 
Nevertheless, according to the standard cosmological paradigm, once this bias due to local motion is accounted for, the cosmological principle is expected to hold.
The near-uniform temperature map of the cosmic microwave background (CMB) strongly supports this principle~\citep{Planck:2019evm}, as do other studies using diverse probes, such as galaxies~\citep[\eg.][]{Marinoni2012JCAP, Alonso2015MNRAS, Sarkar2019MNRAS}, supernovae (SNe)~\citep{Gupta2010MNRAS}, \acp{grb}~\citep[\eg.][]{Meegan1992Natur, Ripa2017ApJ}, and \ac{gw} sources~\citep[\eg.][]{Essick:2022slj,Payne:2020pmc,CalderonBustillo:2024akj}.

Nonetheless, both homogeneity and isotropy remain regularly questioned.
For isotropy, from the \enquote{Axis of Evil} of~\citet{Land:2005ad} to~\citet{Yeung:2022smn}, the principle is often re-investigated.
Many attempts have been discussed, such as in the reviews \citet{Zhao:2015apu,Perivolaropoulos:2021jda,Aluri:2022hzs}.
Several studies reported mild tensions with isotropy.
For example,~\citet{Colin2019A&A} identified a bulk flow inducing a dipolar modulation of the inferred acceleration of expansion rate based on Type Ia SNe data.
Similar hints have been reported in the CMB, where large-scale anomalies suggest potential deviations from statistical isotropy~\citep{Axelsson:2013mva, Schwarz2016CQGra, Yeung:2022smn}.
From the perspectives of \acp{grb},~\citet{Tarnopolski2017MNRAS} showed that statistical tests applied to the Fermi/GBM catalog indicate that short-duration \acp{grb} (sGRBs) exhibit a statistically anisotropic distribution on the sky.
However, other recent analyses of certain datasets revealed dipole amplitudes exceeding predictions from our local motion alone, suggesting a possible \enquote{dipole anomaly} that may indicate departures from perfect isotropy or systematic effects~\citep{Secrest_2021,Secrest_2022,Land-Strykowski:2025gkz,Rasouli:2025yfv,Secrest:2025wyu,kinematic_dipole_4}.
As \citet{Migkas:2016krt} points out, \enquote{precision cosmology may have outgrown the \ac{flrw} paradigm, an extremely pragmatic but non-fundamental symmetry assumption}.

These tensions motivate independent tests using new or complementary observational tools. 
Independent probes provide complementary and reinforcing constraints on the isotropy hypothesis, strengthening confidence in the theoretical framework.
\ac{gw} observations have recently emerged as a powerful and independent probe of large-scale isotropy~\citep{Essick:2022slj, Kashyap_2023}.
Searches for anisotropy in the stochastic \ac{gw} background~\citep{NANOGrav:2023tcn, LIGOScientific:2025bgj} and analyses based on the angular distribution of \ac{gw} sources~\citep{Zheng:2023ezi,Stiskalek:2020wbj,Banagiri:2020kqd} offer new avenues to test isotropy beyond previous probes.
Unlike electromagnetic observations, \ac{gw} detections are not affected by selection effects related to dust, intervening matter, or source variability. 
They are affected however by beam pattern functions, which can be accounted for with the use of synthetic data~\cite{Zheng:2023ezi}. 

Several works have introduced tools to build new probes of this cosmological principle and of the isotropic hypothesis.
For example, \citet{Zheng:2023ezi} tested for anisotropies via \ac{gw} source localization.
It decomposed the \ac{gw} sky localization map (skymap) into spherical modes and compared the result to the expected mode decomposition if \acp{gw} events were isotropically distributed.
This reference decomposition was obtained from synthetic skymaps constructed to be isotropically distributed.
The study used the \ac{gwtc} then available, GWTC-3~\citep{KAGRA:2021vkt}, and restricted itself to three-detector events, totaling 34 events.
The rapid expansion of the \ac{gwtc} now allows revisiting such isotropy tests of individual source populations.
The public release of GWTC-4.0~\citep{ligo2025_gwtc4}, where the upgraded LIGO, Virgo, and KAGRA detectors~\cite{LVK_DET_1,LVK_DET_2,LVK_DET_3,LVK_DET_4,LVK_DET_5,LVK_DET_6} participated, brings the total of observed compact binary merger events to more than 200, and improves sky coverage and statistical power, compared to earlier releases.
The same approach can also be expanded to two-detector events and applied to other sources, such as \acp{grb}.

Here we test the isotropy of the local Universe using the angular distribution of individual transient sources, specifically \acp{gw} and \acp{grb}.
In particular, we use the most recent observational datasets, including \ac{gw} data from the GWTC-4.0 and \ac{grb} data from GRBWeb~\citep{GRBweb, GRBweb2}.
Adopting a methodological framework similar to~\citet{Zheng:2023ezi}, we project source locations onto the sky and decompose the resulting distributions using spherical harmonics.
This approach enables a systematic search for low-order anisotropic modes, such as the dipole moment, and allows us to assess their consistency with the hypothesis of an isotropic distribution.
Therefore, we report the updated cosmological anisotropy results using the latest datasets,
and the first cross-correlation results between \ac{gw} and \ac{grb} catalogs.

\section{Data and Samples}
\label{sec:data}
\begin{figure}
    \centering
    \includegraphics[width=\linewidth]{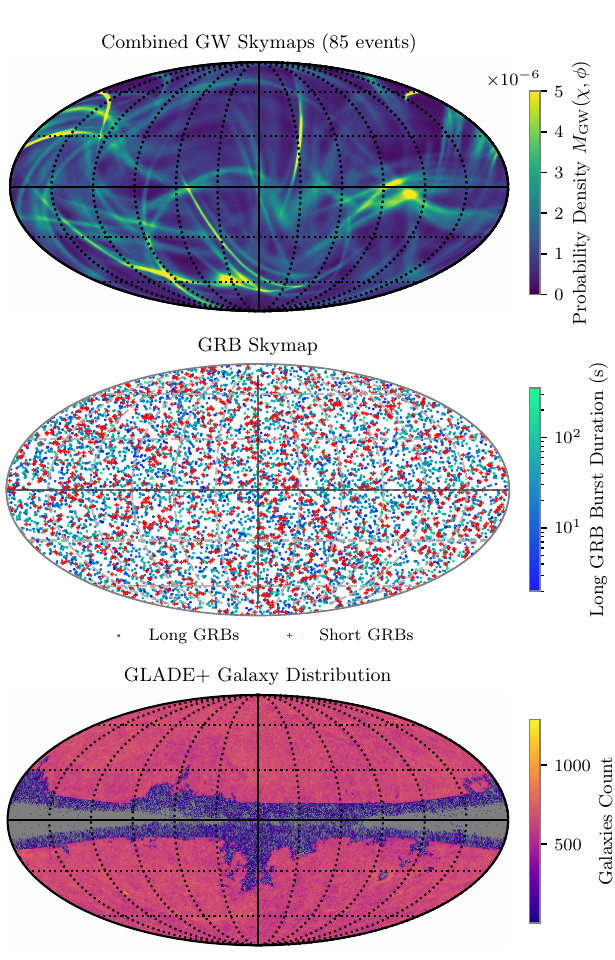}
    \caption{
        Skymap of the new events from GWTC-4.0 (O4a) \cite{ligo2025_gwtc4} \ac{gw} events (top),
        GRBWeb~\citep{GRBweb,GRBweb2} \ac{grb} events since 1991 (middle), and GLADE+ galaxies (bottom) in Galactic coordinates.
        Short \acp{grb} designates burst lasting 2 seconds or less, and the rest are all long \acp{grb}.
        The skymaps and parameter estimation data are used in tests for cosmological anisotropies.
        The GLADE+ galaxy catalog is shown for comparison only; if selection effects are ignored, the distribution appears qualitatively uniform, though the presence of selection effects introduces deviations from isotropy.
    }
    \label{fig:raw_data}
\end{figure}

For \acp{gw}, we include only the new events introduced in GWTC-4.0~\citep{ligo2025_gwtc4}, during O4a, and parameter estimation samples from the \ac{lvk} public release~\citep{LVK2025gwtc4pezenodo}.
As for the \acp{grb}, they are retrieved from GRBWeb~\citep{GRBweb, GRBweb2}, which includes all known \acp{grb} since 1991.
GRBWeb incorporates catalogs such as Fermi-GBM~\citep{Poolakkil2021ApJ_fermi}, SWIFT-BAT \ac{grb} catalogs~\citep{Lien2016ApJ_swiftBAT}, and \acp{grb} reported in the General Coordinates Network (GCN) Circulars~\citep{nasa_grb_circular}.
\Acp{grb} are subdivided into short and long bursts, where short bursts have a duration of less than \qty{2}{\second} ~\citep{Identify_sGRBs, 1984Natur.308..434N}.
Any tests for correlations use the congregated skymaps (\cref{fig:raw_data}).

\begin{figure}
    \centering
    \includegraphics[width=\linewidth]{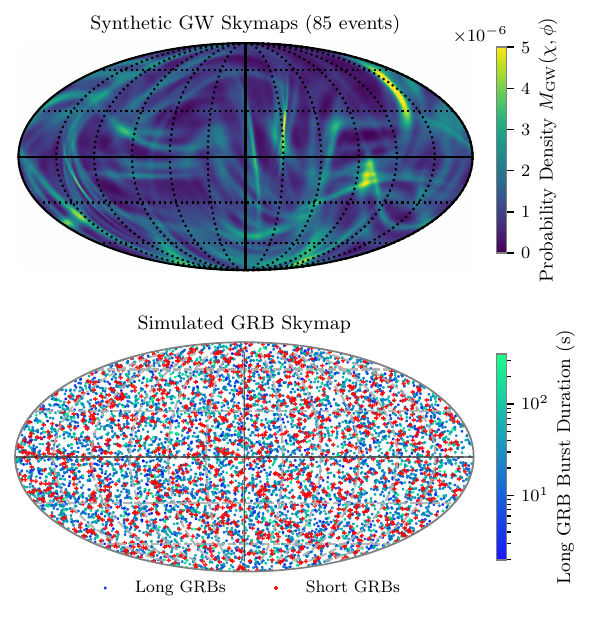}
    \caption{
        Example synthetic skymap of 85 new \ac{gw} events as introduced with GWTC-4.0 (top) and GRBWeb \ac{grb} events (bottom). %
        coordinates for comparison with the raw data.
        The synthetic data is based on the spatial isotropy assumptions and other parameters sampled from the above catalog.
        We use this synthetic data to
        (1) test the methodology against mock data and
        (2) quantify the expected statistical variation in the tests using angular power spectra and the two-point correlation function.
    }
    \label{fig:synthetic_data}
\end{figure}

Synthetic counterpart data for previous \ac{gw} and \ac{grb} data are created to
(1) 
test the methodology against mock data, and
(2) 
quantify expected statistical variation in the tests using angular power spectra and the two-point correlation function (see~\cref{fig:synthetic_data}).

We generate the synthetic \ac{gw} skymaps using \texttt{Bayestar}~\citep{singer2016rapid}, assuming a detector network with LIGO Hanford and Livingston detectors (HL) operating at the averaged O4a sensitivity from~\citet{GWTC4_Sensitivity}.
We produced 7554 synthetic skymaps, enabling 88 O4a-like catalogs with 85 events each. 
The \ac{gw} sources are \ac{bbh} distributed according to the star formation rate~\citep{ligo2025_gwtc4}, with redshift ranges from 0.2 to 1. 
The black hole masses are drawn from the median component mass distribution from GWTC-3~\citep{GWTC3_Population}; with spins distribute isotropically. For the \ac{snr} computation, we adopt IMRPhenomPv2 as the waveform approximant.

For the \acp{grb}, we simulate isotropic sky distribution, with durations from observed data.
We use point estimates for the simulation, as localization uncertainties are arcseconds and are negligible for our analysis.
This assumes that the simulated \ac{grb} data is isotropic, which ignores for example dust extinction, which however is a small effect, and is further confirmed by our results in a later section.
For the duration of \acp{grb}, we fit a bimodal distribution to that of the distribution given by the observed events~\citep{Identify_sGRBs}.
Any possible clustering has been neglected, as the effect is expected to be small (especially taking the redshift into account).

Simulating synthetic data provides insight into the limitations of observational datasets, such as finite sample size effects.
For an isotropic angular distribution, the expected two-point correlation function is flat, serving as a benchmark for comparison.
This simulation approach establishes expectations for isotropically distributed \ac{gw} sources.
The simulated sources span a sufficiently long time such that any selection effects induced by the uneven detector sensitivity on the sky are therefore properly included in the skymaps intrinsically.
Additionally, the sky localization of \ac{gw} events critically depends on the network \ac{snr}, the threshold of which is set to be \num{8}. The loudness of each detection depends on the number and orientation of detectors, and the source’s distance from Earth, introducing intrinsic scales in skymap fluctuations that are difficult to model analytically. 

\section{Methods}
\label{sec:methods}

To test for anisotropy, we first stack all \ac{gw} and \ac{grb} data into a probability distribution function map $M_\text{X}(\chi,\phi)$, where $\text{X}$ labels either the \ac{gw} or \ac{grb} catalog, and decompose the stacked skymap into spherical harmonics:
\begin{equation} \label{eq:SH_decomposition}
    M_\text{X}( \chi, \phi )
        = \sum_{\ell = 0}^{\ell_{\rm max}} \sum_{m = -\ell}^{+\ell}
            {\beta^\text{X}_{\ell m}}
            Y_{\ell m}( \chi, \phi ) \,,
\end{equation}
where $Y_{\ell m}( \chi, \phi )$ are the spherical harmonics\footnote{
    We express the spherical harmonic expansion in colatitude $\chi=\pi/2-\delta$ and azimuthal angle $\phi=\alpha$ instead of right ascension $\alpha$ and declination $\delta$, because $Y_{\ell m}$ and the addition theorem are defined with the polar angle measured from the pole ($\chi=\pi/2-\delta$) instead of declination $\delta$.
}, and $\ell_{\rm max}$ is set to 128.

\begin{figure*}
    \centering
    \includegraphics[width=\textwidth]{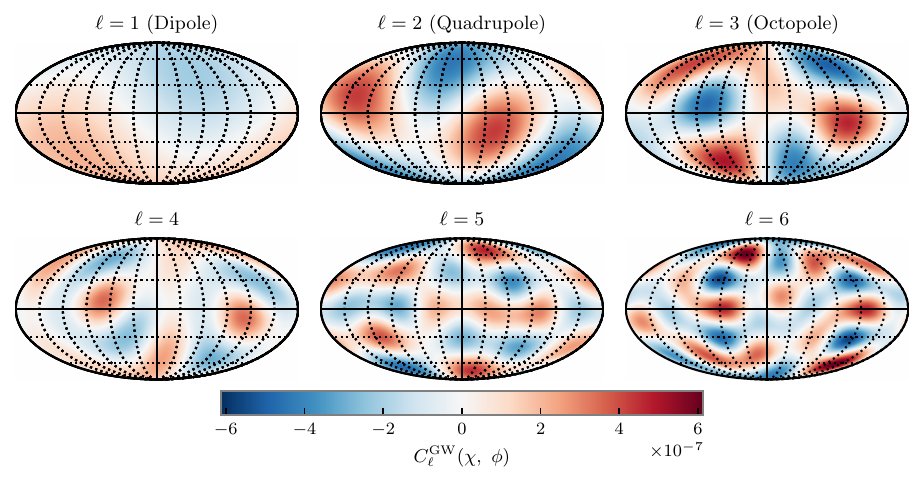}
    \caption{
        First modes of the harmonics decomposition of the cumulative \ac{gw} skymap.
        Each panel shows the contribution from a single multipole moment $\ell$, with all azimuthal orders $m$ summed from $-\ell$ to $\ell$.
        Summing over $m$ yields the total power at each angular scale (approximately $\ang{180}/\ell$), removes dependence on the coordinate system, and reduces statistical noise (\cref{eq:power_spectrum}).
        Low $\ell$ corresponds to large-scale structures across the sky, while higher $\ell$ probes smaller angular scales.
        A deviation in a certain angular multipole moment $\ell$ from the reference isotropic case reflects anisotropy at the corresponding angular scale.
    }
    \label{fig:spherical_decomp}
\end{figure*}

\begin{figure*}
    \centering
    \includegraphics[width=\textwidth]{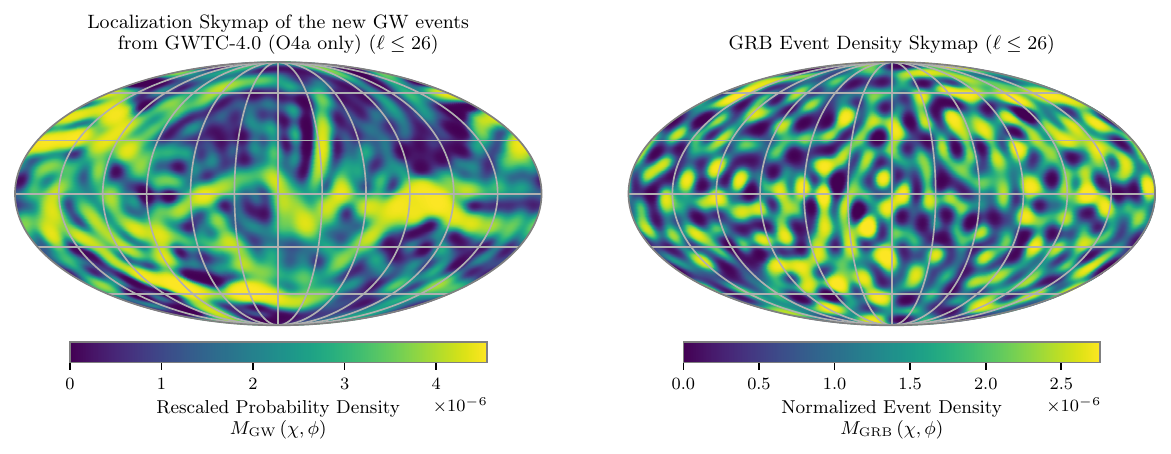}
    \caption{
        Reconstruction from the 85 new events of GWTC-4.0 (left) and \ac{grb} (right) event density skymaps from their harmonics decomposition.
        The \ac{grb} skymap shows the raw data expansion in spherical harmonics, cut at the 26\textsuperscript{th} order in multipoles to match the \enquote{GW} skymap resolution~\citep{Zheng:2023ezi}.
        Both skymaps appear visually isotropic.
        Individual \ac{gw} sky localisations have more significant uncertainties, as most new GWTC-4.0 events (O4a) were detected with two detectors only.
        Such individual sky localisations add power at smaller angular multipole moments $\ell$.
    }
    \label{fig:obs_GRB_skymap_blurred}
\end{figure*}

To measure how strongly any catalog (\eg. \acp{gw} with \acp{gw}) is clumped on a given angular scale ($\sim \ang{180}/\ell$), we compute the angular power spectrum of any catalog of skymaps via
\begin{equation} \label{eq:power_spectrum}
    C_\ell^\text{X} = \frac{1}{2 \ell + 1} \sum_{m} \beta^\text{X}_{\ell m} \bar{\beta}^\text{X}_{\ell m}\,.
\end{equation}
Summing azimuthal orders removes coordinate dependence and reduces statistical noise.
We use these angular (auto) power spectra to test for cosmological anisotropies by comparing them against a fiducial distribution (\cref{sec:data}).
\Cref{fig:spherical_decomp} gives the first few modes of the \ac{gw} skymap decomposition, while \cref{fig:obs_GRB_skymap_blurred} shows reconstructed skymaps from the decomposition.\footnote{The probabilistic map for the GRBs are built by decomposing the spatial distribution of GRBs in spherical harmonics. Then, the decomposition is cut at the 26\textsuperscript{th} order in multipoles and normalized, to match the resolution of GW skymap.}

Once decomposed into spherical harmonics, the angular cross-power spectrum can be written as
\begin{equation} \label{eq:cross_power_spectrum}
    C_\ell^{\text{X}\times \text{Y}} = \frac{1}{2\ell+1} \sum_{m} \beta^\text{X}_{\ell m} \bar{\beta}^\text{Y}_{\ell m} \,,
\end{equation}
where $\text{X}$ and $\text{Y}$ designate two types of data (\eg. \acp{gw} and \acp{grb}).
The goal is to compare the power spectrum of a dataset with the synthetic versions of that dataset, built under the hypothesis of isotropy.
Note that this approach does not infer the preferred direction, but a disagreement suggests the Universe has some anisotropic component affecting \acp{gw}, and/or \acp{grb}. 

To measure how likely two events from different catalogs (\ac{gw} + \ac{grb}) are to be separated by angle $\theta$,
we compute the angular cross-correlation function
\begin{equation} \label{eq:angular_cross_correlation}
    C_{\text{X}\times \text{Y}}\left( \theta \right)
        = \frac{1}{4 \pi}
        \sum_{\ell, m}
        W_\ell(\theta)
            \beta^\text{X}_{\ell m}
            \bar{\beta}^\text{Y}_{\ell m}
            P_\ell\left( \cos{\theta} \right)\,,
\end{equation}
where $P_\ell(\cos \theta)$ is the Legendre polynomial of order $\ell$ used to transform between the harmonic and angular space.
As in \cite{Zheng:2023ezi}, the summation over $\ell$ is up to $\ell = 128$, and $W_\ell$ is the window function that suppresses contributions from modes with $\ell > 26$, as in \cite{Zheng:2023ezi}.
We also compute the angular auto-correlation function for each catalog by
\begin{equation} \label{eq:angular_auto_correlation}
    C_\text{X} (\theta)
        \equiv
        C_{\text{X}\times \text{X}}\left( \theta \right)
        = \frac{1}{4\pi} \sum_{\ell,m}
        W_\ell(\theta)
        \beta^\text{X}_{\ell m} \bar{\beta}^\text{X}_{\ell m} \, P_\ell(\cos\theta)\;.
\end{equation}
The auto-correlation functions are less susceptible to observational systematics than the angular power spectra.
For \acp{gw}, it is because they are less sensitive to sky localization uncertainties (which add power at high angular mode $\ell$).
The expectation from an isotropic distribution is a flat autocorrelation function. 
This remains the case even if the merger-rate density is changed, though it changes the cut-off scale at which the angular autocorrelation is no longer resolved.

\section{Results}
\label{sec:results}
\begin{figure}
    \centering
    \includegraphics[width=\linewidth]{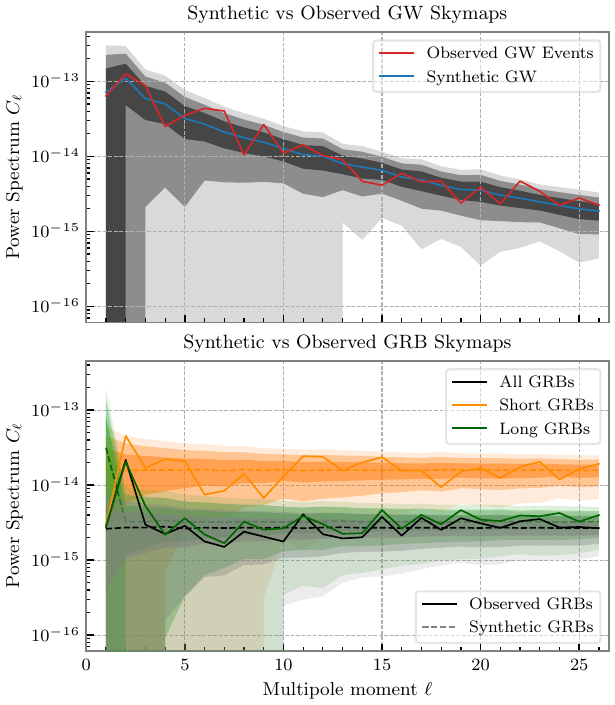}
    \caption{
        The observed \ac{gw} (top) and \ac{grb} (bottom angular power spectrum for $1 \leqslant \ell \leqslant 26$).
        In both the \ac{gw} and \ac{grb}, the synthetic data (median in blue; \qtylist{1;2;3}{\sigmaunit} uncertainties in gray) contains the observed data (red/orange/green).
        Neither dataset shows deviations from anisotropy.
        There are slight discrepancies between the datasets, with less angular power in \acp{gw} at larger angular multipole $\ell$ corresponding to smaller scales, and more uniformly spread power in \acp{grb}.
        This discrepancy is expected from the greater uncertainty in sky localizations of \acp{gw}, where each map may span a larger sky area compared to the \ac{grb} sky localizations, depending on the number of detectors online at a given time and the signal-to-noise ratio.
    }
    \label{fig:ang_pow_spec}
\end{figure}

To quantify the power at different angular scales, we compute the observed angular power spectrum (\cref{eq:power_spectrum}) for the 85 new events from the \ac{lvk} GWTC-4.0~\citep{ligo2025_gwtc4} and for the entire GRBWeb catalog.
Unlike the flat spectra of the \ac{grb} distributions, the \ac{gw} power spectrum displays a strong peak at low multipoles and decays with increasing $\ell$.
The prominent peak at small $\ell$ results from the larger spread of \ac{gw} skymaps than the point-like \ac{grb} sources (\cref{fig:ang_pow_spec}).
While the \ac{grb} spectra reflect the spatial distribution of point sources, the \ac{gw} spectra incorporates the limitation in precision of the \ac{gw} sky localisation.
In both cases, the observed power spectra fall within the uncertainty bands of the isotropic synthetic dataset, suggesting no significant deviation from the cosmological principle.

\begin{figure}
    \centering
    \includegraphics[width=\linewidth]{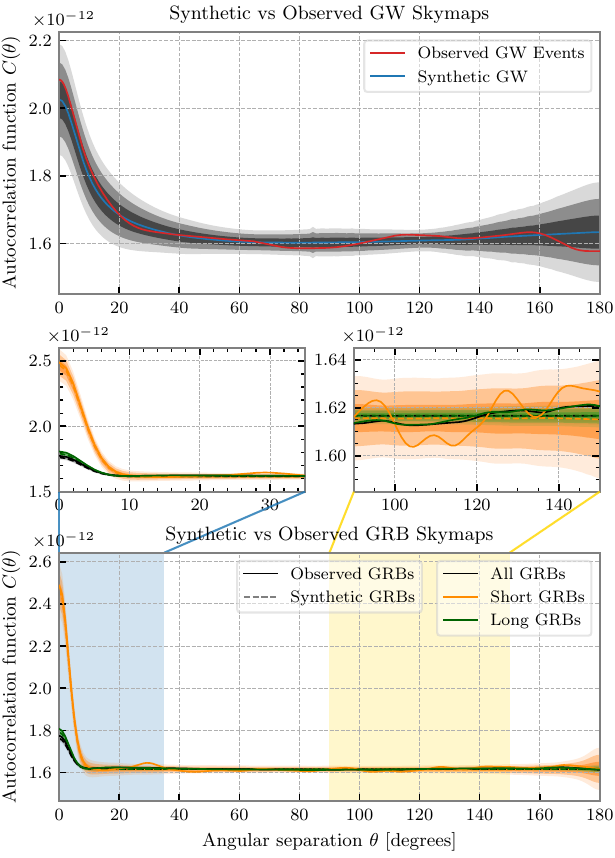}
    \caption{
        The auto-correlation function between the observed \ac{gw} (top) and \ac{grb} (bottom) data against the fiducial model built under the isotropic assumption (shaded \qtylist{1;2;3}{\sigmaunit} regions).
        Apart from angles \ang{< 20}, the close alignment indicates consistency with the isotropic distribution.
        The auto-correlation function of \ac{grb} events, with spatial positions approximated to $\ell\leqslant 26$ to match the resolution of \acp{gw} positions. The light-colored bands represent deviations from the synthetic isotropic \acp{grb} from \qtyrange{1}{3}{\sigmaunit}.
        At smaller scales, near to \ang{0}, however, the spectra diverge from the expectation of isotropy (flatness).
        This divergence is likely driven by the finite number of observations and, in the case of \acp{gw}, the limitations in the sky localisation (some visual anisotropy is apparent even in raw and synthetic data).
        The auto-correlation functions of all and long \acp{grb} are within a \qty{3}{\sigmaunit} limit, while that of short \acp{grb} deviates more and occasionally exceeds a \qty{3}{\sigmaunit} limit.
        However, because the observed short \acp{grb} are only $1/6$ of all \acp{grb}, and are accounted for in the simulations; anisotropy cannot be confirmed from this result alone.
    }
    \label{fig:GW_GRB_Auto_Correlation}
\end{figure}

Under the hypothesis of isotropy, the angular auto-correlation is expected to remain constant across all scales.
The auto-correlation angular spectra for both \acp{gw} and \acp{grb} exhibit this flat behavior between \ang{60} to \ang{180} (\cref{fig:GW_GRB_Auto_Correlation}).
At smaller scales, near to \ang{0}, however, the spectra diverge from the expectation of isotropy.
This divergence is likely driven by the finite number of observations and, in the case of \acp{gw}, the limitations in the sky localisation (some visual anisotropy is apparent even in raw and synthetic data; \cref{fig:raw_data,fig:synthetic_data}).

\begin{figure}
    \centering
    \includegraphics[width=\columnwidth]{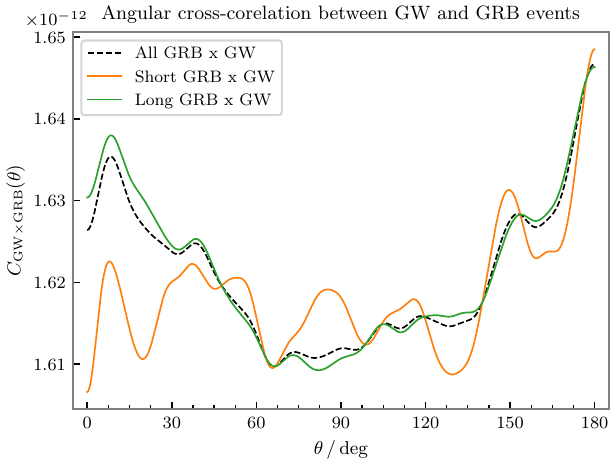}
    \caption{
        The angular cross-correlation function between \ac{grb} and \ac{gw} events, defined in \cref{eq:angular_cross_correlation}.
        The dashed curve shows the cross-correlation for all \acp{grb}; whereas the orange and green solid curves show respectively the cases for short and long \acp{grb}.
        Generally, the short \acp{grb} show larger fluctuations and follow less closely with the overall distribution (black), for the reason explained in \cref{fig:GW_GRB_Auto_Correlation}.
        Nevertheless, all three curves hover close to \num{1.62e-12}, and fluctuations in all three cases are relatively small, at the order of $10^{-14}$.
        These suggest that the curves are mostly flat, and there is no strong correlation between \ac{grb} and \ac{gw}, at any angular scales.
    }
    \label{fig:GW_GRB_Cross_Correlation}
\end{figure}

Cross-correlation fluctuations occur at scales consistent with the auto-correlation spectra, though they exhibit a reduced variance ($\mathord{\sim}\num{1.4e-12}$) compared to the full GRB auto-correlation (\cref{fig:GW_GRB_Cross_Correlation}).
The larger deviations observed in the short \ac{gw}-\ac{grb} cross-correlation arise from similar resolution constraints as in the auto-correlation, where the smaller number of short \ac{grb} events leads to larger statistical fluctuations.
The low cross-correlation values suggest only a weak spatial correlation between these messengers.
While current results show no signature of anisotropy, a rejection of the isotropic-but-correlated scenario requires comparison with synthetic datasets in follow-up analyses.

\section{Discussion and conclusions}
\label{sec:conclusions}

In this work, we used the latest datasets for new \ac{gw} events and \acp{grb} from the GWTC-4.0 (O4a only) and GRBWeb catalogs and carried out two tests.
Firstly, we used the method by \citet{Zheng:2023ezi} to search for evidence of anisotropy.
We generated synthetic isotropically distributed data for comparison.
We decomposed the cumulative skymap of all recorded events into spherical harmonics and obtained the angular power spectrum.
This analysis revealed no significant sign of anisotropy (\cref{fig:ang_pow_spec,fig:GW_GRB_Auto_Correlation}), upholding the cosmological principle's validity.

Secondly, the observational power spectra are used to look for cross-correlations between \acp{gw} source localization and \acp{grb} (both separated into short and long bursts and aggregated).
No apparent correlation stands out (\cref{fig:GW_GRB_Cross_Correlation}), though a more thorough analysis could find a positive result.

This work concludes that the cosmological principle's validity remains plausible.
More thorough analyses of these datasets, \eg. carefully including subtle selection functions, could provide valuable complementary information about our Universe and the mathematical foundations of its description.

\vskip 5.8mm plus 1mm minus 1mm
\vskip1sp
This work comes, in part, in the continuity of the \href{https://gw.phy.cuhk.edu.hk/activities/gw-hands-on-workshop-2025/}{Gravitational-Wave Hands-on Workshop 2025}, held at The Chinese University of Hong Kong between the 27\textsuperscript{th} of November and the 1\textsuperscript{st} of December.
The code used to make the figures in this work is freely available on \href{https://github.com/SSL32081/Testing-anisotropy-with-GW-and-GRB}{GitHub} (\texttt{SSL32081/Testing-anisotropy-with-GW-and-GRB}).
BHYC, OAH, SJ, PKKL, SCL, SHWL, ML, PM, TCKN, HX and BHLN acknowledge support by grants from the Research Grants Council of Hong Kong (Project No. CUHK 14304622, 14307923, 14307724, AoE/P-404/18, and 14300223), the start-up grant from The Chinese University of Hong Kong, the Direct Grant for Research and Research Fellowship Schemes from the Research Committee of The Chinese University of Hong Kong.
TCKN acknowledges support by the research program of the Netherlands Organization (NWO).
GS acknowledges support by the National Research Foundation of Korea (NRF) grant funded by the Korea government (MSIT) (No. RS-2024-00394623).
DCC acknowledges support by the Department of Science and Technology - Science Education Institute of the Philippines through the ASTHRDP-NSC scholarship.
A subset of the results in this paper has been generated using the healpy and HEALPix packages~\citep{Gorski:2004by, Zonca:2019vzt}.
This research has made use of data or software obtained from the Gravitational Wave Open Science Center (gwosc.org), a service of the LIGO Scientific Collaboration, the Virgo Collaboration, and KAGRA.
This material is based upon work supported by NSF's LIGO Laboratory which is a major facility fully funded by the National Science Foundation, as well as the Science and Technology Facilities Council (STFC) of the United Kingdom, the Max-Planck-Society (MPS), and the State of Niedersachsen/Germany for support of the construction of Advanced LIGO and construction and operation of the GEO600 detector.
Additional support for Advanced LIGO was provided by the Australian Research Council.
Virgo is funded, through the European Gravitational Observatory (EGO), by the French Centre National de Recherche Scientifique (CNRS), the Italian Istituto Nazionale di Fisica Nucleare (INFN) and the Dutch Nikhef, with contributions by institutions from Belgium, Germany, Greece, Hungary, Ireland, Japan, Monaco, Poland, Portugal, Spain.
KAGRA is supported by Ministry of Education, Culture, Sports, Science and Technology (MEXT), Japan Society for the Promotion of Science (JSPS) in Japan; National Research Foundation (NRF) and Ministry of Science and ICT (MSIT) in Korea; Academia Sinica (AS) and National Science and Technology Council (NSTC) in Taiwan.
The authors are grateful for computational resources provided by the LIGO Laboratory and supported by National Science Foundation Grants PHY-0757058 and PHY-0823459.
We acknowledge the use of computing facilities supported by grants from the Croucher Innovation Award from the Croucher Foundation, Hong Kong.
This manuscript has LIGO-DCC number P2500805.

\bibliographystyle{aasjournal}   %
\bibliography{references}
\end{document}